\documentclass[review,12pt]{elsarticle}
\usepackage{graphics}
\usepackage{amssymb}
\usepackage{bm}
\usepackage{soul}
\usepackage{color}
\journal{Annals of Physics}
\begin{document}
\begin{frontmatter}

\title{Phase space quantum-classical hybrid model}
\author{Gerardo Garc\'ia }
\author{Laura Ares}
\author{Alfredo Luis \corref{ca}}
\ead{alluis@ucm.es}
\cortext[ca]{Corresponding author}

\address{Departamento de \'{O}ptica, Facultad de Ciencias F\'{\i}sicas, Universidad 
Complutense, 28040 Madrid, Spain}

\begin{abstract}
In this work we provide a complete model of semiclassical theories by including back-reaction and correlation into the picture. We specially aim at the interaction between light and a two-level atom, and we also illustrate it  via the coupling of two harmonic oscillators. Quantum and classical systems are treated on the same grounds via the Wigner-Weyl phase-space correspondence of the quantum theory. We show that this model provides a suitable mixture of the quantum and classical degrees of freedom, including the fact that the evolution transfers nonclassical features to the classical subsystem and nonquantum behavior to the quantum subsystem. In that sense, we can no longer distinguish between classical and quantum variables and we need to talk about a hybrid model.
\end{abstract}


\end{frontmatter}

\section{Introduction}

Semiclassical approaches play several roles with respect to a fully quantum theory \cite{MWSZ1,MWSZ2}.  On the one hand, they are useful approximations to solve problems that otherwise would be unnecessarily complex.  On the other hand, they are suitable tests regarding the necessity of the quantum theory. This is to say, we may consider that a phenomenon supports the quantum theory provided that it cannot be explained within any  classical or semiclassical theory. We can invoke for example the case of the photoelectric effect that would not  be a proof of the quantum nature of radiation as far as it admits a semiclassical explanation \cite{LS691,LS692}. Beyond light-matter interaction, these hybrid quantum-classical theories are examined in many works in different areas, including quantum gravity where the inconsistency of simple semiclassical models suggests that a full quantum theory or a hybrid approach to the problem is needed. This is also a suitable framework to address the quantum to classical  transition, or even the quantum measurement problem, where the issue of consistency is also carefully examined \cite{Hy1,LG,HTE1,HTE2,HTE3,HR,BCLIR,CS}. Specifically we focus on the light-matter interaction, exemplified as the interaction of a two-level atom and a one-mode field. \\

Semiclassical theories usually involve some kind of approximation that neglects the possible influence of the quantum subsystem on the classical one. Even if they consider it, they usually are included in some kind of mean field theory such as the self consistent Hartree-Fock approximation for describing atomic structure. For instance, we can consider the paradigmatic case of Bose-Einstein condensates, which are usually described in terms of a condensed part and a non-condensed one that constitutes the elementary excitations of the system, whose back reaction on the condensed part is usually disregarded or it is included in terms of the so called Hartree-Fock-Bogoliubov approximation \cite{GR96}. This kind of approximation reduces the back reaction on the condensed part by an effective mean field term from the beginning, and thus a full dynamical evolution of the system is not taken into account. The light matter interaction theory is not different from other semiclassical approaches and it usually neglects the back-reaction of the atom on the field as well as atom-field correlations. \\

In this work, we propose to go further completing semiclassical models by including both back-reaction and correlations. The way we propose to combine classical and quantum systems is provided by the phase-space representations of quantum  physics \cite{WC1,WC2,WC3,WC4,WC5,VG891,VG892,VG893,BT,AMT}. Among them, the Wigner-Weyl correspondence has the enormous advantage that in rather interesting scenarios it propagates as it were a classical phase-space distribution \cite{LG}. In such case the classical and quantum variables propagate exactly in the same form. The main properties of this phase-space representation are summarized in Appendix A. For completeness in Appendix C we consider also the simpler case of two coupled harmonic oscillators: one quantum and the other classical. \\

Such a more complete model presented in this work should be able to improve the power of the semiclassical approximations and properly investigate its own limits. For example, we can examine whether the differentiation between classical and quantum variables is still meaningful. This is because evolution may transfer quantum features to the classical system and the other way round, a sign that we can not distinguish between classical and quantum subsystems and we are forced to talk about hybrid subsystems, since they can display classical and quantum features. This model can be interesting to better know the region where the distinction between classical and quantum degrees of freedom can be made consistently since dynamical evolution does not mix them.  \\

In that sense, our proposal goes further than the usual mean field approach that decouples effective classical degrees of freedom and studies quantum fluctuations propagating over that fixed background which are the classical degrees of freedom. Because of that, it can be seen as a semiclassical approximation with well-differentiated classical and quantum degrees of freedom. Even if the back reaction of that fluctuations is taken into account on the background, it is just done at a certain degree of accuracy by including correlations up to the desired order and thus, from the beginning, one neglects the possibility of a mixture between those classical and quantum degrees of freedom (hybridisation). Usually, including higher order correlations becomes hard and thus, in practice, just the first few orders are taken into account. Our model fills the gap of an analysis of the full dynamics of a classical and quantum subsystems coupled, not relying on perturbative expansions or assumptions on the importance of correlations and allowing that hybridisation between the classical and quantum degrees of freedom. \\

It is worth comparing our formalism with other hybrid models that also include nontrivial dynamics of the classical system and back-reaction which have been discussed in the literature so far, see Refs. \cite{Hy1,LG,HTE1,HTE2,HTE3,HR,BCLIR,CS} and references therein. The key point is that our formalism is based on a mathematical structure common to classical and quantum mechanics: the phase space. Moreover, the evolution form proposed here grants a rather relevant point: in the absence of interaction each subsystem preserves its natural evolution equations. These are worthwhile features of our model when contrasted to other approaches. More specifically the approach presented in this work involves no translation of classical mechanics into the language of Hilbert spaces as in Refs. \cite{Hy1,LG} where the classical nature is then retrieved by additional hypotheses, such as normal ordering and coherent states. In other approaches it is the quantum system which is formulated in a classical-like form as in Refs. \cite{HTE1,HTE2,HTE3} but in terms of variables that only exist in the quantum domain, instead of the physical phase-space variables invoked in our formulation. Moreover, no unobservable variables or additional structures must be invoked to maintain the commutation of classical observables in Hilbert space nor special quantum-classical bracket has to be defined, neither is necessary to generalize Poisson and commutator brackets as in Refs. \cite{HR,BCLIR,CS}. 

In this regard, our approach might suggest that coupling a classical system to a quantum one may be possible in a consistent way.  This points to the existence of a general framework where classical and quantum features can be displayed by both subsystems as a consequence of the dynamical mixing induced by interactions. In that sense, this model provides a different framework from the usual semiclassical picture and a fully quantum one, suggesting that novel features can emerge from this kind of hybrid theories \cite{Hy1,LG}. \\

\section{Model}

Let us focus on the semiclassical theory of matter-light interaction. More specifically:  the quantum system is a two-level atom while the classical system is a one-mode field. Their coupling is one of the most fruitful models in quantum optics \cite{AE87}.

The optimal arena to mix quantum and classical degrees of freedom is the Wigner-Weyl phase-space scenario, where physical states are represented by functions $W (z)$ on the corresponding phase space of the system parametrized by some set of coordinates $z$. Within this picture the evolution of the system is just given by the Liouville equation in terms of the corresponding Poisson brackets
\begin{equation}
\dot{W} (z)=\{H (z),W (z) \}. 
\end{equation}
This is to say that the evolution of $W(z)$ is given by the classical-like evolution of the phase-space coordinates $z$ via the known relation:
\begin{equation}
\label{evol}
W_t \left [ z (t) \right ]  = W_0 (z) \rightarrow W_t (z) = W_0 \left [  z(-t) \right ] ,
\end{equation}
It is worth noting that this is valid for any Hamiltonian $H(z)$. 
The solution to this evolution equation is particularly simple if the dynamics leads to a linear transformation, i. e., 
\begin{equation}
\label{We}
\mathrm{if} \quad z(t) = M z,  \quad \mathrm{then} \quad W_{U \rho U^\dagger} ( M z )  = W_\rho ( z ),
\end{equation}
where $z(0)=z$ and $U$ is the unitary operator representing the evolution in quantum mechanics.

A key point is that this naturally includes the back action of the atom on the field beyond mean values to propagate the complete distribution without assuming any kind of atom-field factorization \cite{CLS99}. \\
 
In this regard, the most fruitful formulation for a spin-like finite-dimensional system is provided in Ref. \cite{VG891,VG892,VG893} via the SU(2) Wigner function $W_q (\Omega)$ recalled in Appendix A2,  where the corresponding phase space is the Bloch sphere parametrized by the coordinates $\Omega = \{ \theta, \varphi \}$, being $\theta$ and $\varphi$ the polar and azimuthal angles respectively. In this case, the covariance condition guaranteeing classical-like evolution of the Wigner function holds for SU(2) transformations, i. e., rotations on the Bloch sphere. The field is considered as a single mode described via a Wigner-like function $W_c (\alpha)$ where $\alpha = r e^{-i \phi}$ is the field complex-amplitude, i. e., the phase space for the field mode. The combination of SU(2) and Cartesian variables has been studied previously for Galilean particles with spin in Ref. \cite{VG88}. \\

This provides us with a very convenient program: 

\bigskip

1) We specify a legitimate Wigner function $W_q (\Omega )$ for the initial state $\rho$ of the quantum subsystem via the first equation in Eq. (\ref{Wr}). We specify also a {\em bona fide}  phase-space distribution function $W_c (\alpha)$ describing the classical subsystem.  So the initial state of the whole system in the phase-space picture is $W_0 (\alpha, \Omega )= W_c (\alpha) W_q (\Omega )$. 

\bigskip

2) The system evolves in phase space under the classical evolution according to the classical-like form in Eq. (\ref{evol}).

\bigskip

3) Finally, if desired, we may extract the state of the system in the quantum and classical variables. They are separated taking the corresponding marginals 
\begin{equation}
W_t (\alpha ) = \int \textrm{d}  \Omega W_t (\alpha, \Omega) , \quad  W_t (\Omega ) = \int \textrm{d} ^2 \alpha W_t (\alpha, \Omega) ,
\end{equation}
with $\textrm{d} \Omega = \sin \theta \textrm{d} \theta \textrm{d} \varphi$ and $\textrm{d}^2 \alpha = r \textrm{d} r \textrm{d} \phi$. Then we can convert $ W_t (\beta )$ into a Hilbert-space operator via the second relation in Eq. (\ref{Wr}).
 
\section{Non resonant model}

The typical quasi resonant atom-field interaction produces rather cumbersome transformations \cite{CLS99}. In order to illustrate the main ideas behind the model let us consider the limit of strong detuning where the interaction Hamiltonian in a fully quantum picture is of the form
\begin{equation}
\label{Hi}
H_{\rm int} = \chi a^\dagger a \sigma_z ,
\end{equation}
where $a$ is the field-mode complex-amplitude operator, $\chi$ is a coupling constant, and $\sigma_z = | e\rangle \langle e | - | g \rangle \langle g |$ where $| e \rangle$ and $| g \rangle$ are the excited and ground states. This is an interaction term suitable for producing strong nonclassical effects in the form of Schr\"{o}dinger cats for example \cite{AL02} among other beautiful experiments \cite{di1,di2,di3,di4,di5,di6,di7}.

\bigskip

Next we proceed to solve the evolution for the phase-space variables  $\gamma = (\alpha, \Omega)$. The interaction Hamiltonian in Eq. (\ref{Hi}) is the product of the infinitesimal generators of shifts on the field phase and shifts on the azimuthal angle of the Bloch sphere. This holds equally well both in the classical and quantum pictures. If we write the field complex-amplitude as $\alpha = r e^{-i \phi}$ we have, within the interaction picture,  
\begin{equation}
\label{br}
r (t) = r(0), \qquad  \phi(t) = \phi (0) + \sqrt{3} \chi t \cos \theta, 
\end{equation}
where we have expressed $\sigma_z$ as $\sqrt{3}\cos \theta$. The evolution of the atomic variables is 
\begin{equation}
\theta (t) = \theta (0) , \qquad  \varphi(t) = \varphi (0) + 2 \chi r^2 t  .
\end{equation}
We can appreciate in Eq. (\ref{br}) the back-reaction of the atom on the field. This is concentrated on the field phase since in this non resonant scenario the modulus of the amplitude $r$ remains constant. 

\bigskip

Therefore,  the joint Wigner function evolves in the form 
\begin{equation}
\label{completeWt}
W_t (\Omega, \alpha) = W_q (\theta,  \varphi  - 2 \chi r^2 t ) W_c (r,  \phi - \sqrt{3} \chi t \cos \theta) .
\end{equation}
Then we investigate the time evolution of the marginal distributions for the field and the atom by integrating with respect of the atom and field variables, respectively,
\begin{eqnarray}
\label{Wc}
& W_t (\alpha) = \int \textrm{d}  \Omega W_q (\theta,  \varphi  - 2 \chi r^2 t  ) W_c (r,  \phi - \sqrt{3} \chi t \cos \theta), & \nonumber \\
 & & \\
 & W_t (\Omega) = \int \textrm{d} ^2 \alpha W_q (\theta,  \varphi  - 2 \chi r^2 t  ) W_c (r,  \phi - \sqrt{3} \chi t \cos \theta) .& \nonumber
\end{eqnarray}

\section{Standard semiclassical model}

The standard semiclassical model neglects the back-reaction of the atom on the field by disregarding any field evolution other than the free evolution. This means constant field within the interaction picture adopted here. So, there is no point in caring about the field distribution in time and we can focus just on the atomic evolution via its SU(2) Wigner function as:
\begin{equation}
\label{eqq}
W_t (\Omega) = \int \textrm{d} ^2 \alpha W_q (\theta,  \varphi  - 2 \chi r^2 t ) W_c (\alpha)  .
\end{equation} 
Moreover, sometimes a simple mean field approximation is done so evolution is further simplified as 
\begin{equation}
\label{eqq2}
W_t (\Omega) = W_q (\theta,  \varphi  - 2 \chi \langle r^2  \rangle t )  ,
\end{equation} 
where $ \langle r^2  \rangle$ represents the mean intensity of the field. \\

Our model goes further than this standard semiclassical model and includes the back reaction of the atom on the field as it can be seen comparing Eq. (\ref{Wc}) with Eqs. (\ref{eqq}) or (\ref{eqq2}). Moreover, we do not introduce it in terms of some mean field expression; we give the complete evolution of the system in Eq.  (\ref{completeWt}) and thus obtain a non-perturbative evolution of the full system.

\section{Quantum model}

Naturally  it is worth comparing the semiclassical picture with the completely quantum model. Focusing on pure states that factorize at $t=0$ as $| \Psi (0) \rangle = |\psi_a \rangle |\psi_f \rangle$ where $|\psi_a \rangle$ represents the most general atomic pure state
\begin{equation}
 |\psi_a \rangle = c_e | e \rangle + c_g | g \rangle, 
\end{equation}
and $|\psi_f \rangle$ any field state, we have that for any time $t$ the state of the system is 
\begin{equation}
| \Psi (t) \rangle = c_e | e \rangle e^{-i \chi t a^\dagger a}  |\psi_f \rangle + c_g | g \rangle e^{i \chi t a^\dagger a}  |\psi_f \rangle ,
\end{equation}
which is in general an entangled atom-field state. In particular, simpler expressions are obtained if the initial field state is coherent $|\alpha \rangle$, which is usually understood as the most classical field state, leading to 
\begin{equation}
| \Psi (t) \rangle = c_e | e \rangle  |e^{-i \chi t}  \alpha \rangle + c_g | g \rangle  |e^{i \chi t} \alpha \rangle .
\end{equation}

We can appreciate that our semiclassical model includes the shift of the field phase caused by the back reaction of the atom on the field. Naturally, the difference is that in the quantum case only two values for the phase shift are possible, while in a semiclassical picture they are continuously distributed between the two extreme values, recalling the difference between classical and quantum angular-momentum components. \\

Clearly, the phase-space picture will not provide always the exact quantum evolution when the initial field distribution $W_c (\alpha )$, even  is the Wigner function of a legitimate quantum state. In relation with Ref. \cite{VG88} we have that phase space rotations are not part of the Galilean group.

\bigskip

\section{Atom initially in the ground state}

For definiteness, in the following example we will consider the most simple case in which the atomic state is the ground state $|g \rangle$ with
\begin{equation}
\label{Wqgs}
W_q (\Omega) = \frac{1}{4 \pi} \left ( 1- \sqrt{3} \cos \theta \right ) ,
\end{equation}
that takes negative values around the north pole $\theta =0$. As for the field, let us consider two alternative initial states. 

\subsection{Example: initial field state with perfectly defined amplitude}

Let us start with a field with a perfectly defined complex amplitude, this is the Dirac delta functions for amplitude and phase. This is the extreme classical case for the field. Since the behavior of $r$ is trivial let us focus just on the behaviour of the phase, leading to 
\begin{equation}
W_t (\phi) =  \frac{1}{2} \int \textrm{d}  \theta \sin \theta \left ( 1- \sqrt{3} \cos \theta \right ) \delta \left (  \phi - \sqrt{3} \chi t \cos \theta \right ) .
\end{equation}
Using the properties of the delta distribution we obtain the following expression 
\begin{equation}
\label{pf}
W_t (\phi) = \frac{1}{2 \sqrt{3} \chi t} \left( 1 - \frac{\phi}{\chi t} \right),
\end{equation}
for $ \sqrt{3} \chi t \geq \phi \geq - \sqrt{3} \chi t$ and $W_t (\phi) =0$ otherwise. The key point is that this phase distribution $W_t (\phi) $ takes negative values for $\sqrt{3} \chi t \geq \phi > \chi t $, revealing our inability to distinguish between classical and quantum variables and introducing the necessity of talking about hybrid models as we advanced in the introduction.  

\bigskip

For the sake of comparison, in the standard semiclassical model there is no evolution at all for the field, while in the fully quantum picture $ |\Psi (t) \rangle = | g \rangle  |e^{i \chi t} \alpha \rangle$ so we may safely say that the field phase evolves linearly in time as  $\phi (t) = - \chi t$. Beyond these simple comments there is no much point of comparison with the quantum model since there cannot be any quantum state with perfectly defined complex amplitude, as exemplified by Heisenberg uncertainty relations.  This is to say that such initial state for the field would be highly nonquantum.

\bigskip

Let us comment on four points regarding the result of this simple case. (i) $W_t (\phi) $ takes negative values for any $t>0$ no matter how small is $t$. So the appearance of hybrid variables revealed here is not the cumulative effect of maintaining an approximation beyond its natural limits. It is instantaneous, so it relies at the heart of the model and is not a matter of regimes of application. (ii) We emphasize that the atom-field coupling involves no real photon emission nor absorption. (iii) The pathological behaviour reported here holds for every field intensity. This includes the case of fields with large enough intensities that should be unaffected by their coupling to an atom, according to standard formulations of the semiclassical model. Once again, this points to the fact that the hybridisation of our degrees of freedom is of fundamental character and not a matter of regimes of validity. (iv) As a further comparison with standard semiclassical and fully quantum pictures we may compute the first moments of the field phase to get 
\begin{equation}
\langle \phi \rangle = - \chi t , \qquad \Delta \phi = 0. 
\end{equation}
We see that the field phase and the amplitude follow the same quantum evolution. It is really curious how the negativity of $W_t (\phi)$ maintains the initial null variance for the phase at the initial classical field state, at least formally, at the prize of a nonpositive distribution. On the other hand from Eq. (\ref{Wc}) the marginal Wigner function for the atomic state is always the one for the ground state. So our complete semiclassical model is as close as possible to the full quantum model.

\bigskip

In Fig. 1 we can see nontrivial results for the predictions of our extended semiclassical model for the evolution of the correlations $\langle \sigma_z a \rangle - \langle \sigma_z \rangle \langle a \rangle$ for a perfectly defined field of unit amplitude. This correlation vanishes in the standard semiclassical and fully quantum models, while in this extended semiclassical model tends to zero as $1/t$ when $t$ increases. Similarly, in Fig. 2 we plot the evolution of the field complex amplitude $\langle a \rangle$ under the same conditions, decaying to zero in our model (dashed line), as a consequence of the back-reaction of the atom on the field phase,  while the standard semiclassical and fully quantum models predict a constant-amplitude oscillation (solid line). 
\begin{figure}
\begin{center}
\includegraphics[width=8cm]{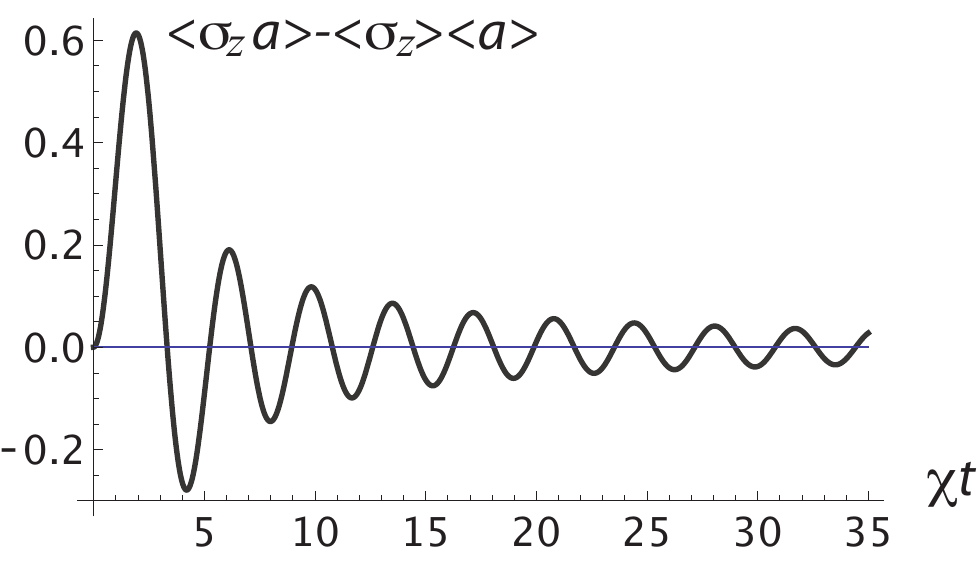}
\end{center}
\caption{Evolution of the correlations $\langle \sigma_z a \rangle - \langle \sigma_z \rangle \langle a \rangle$ as a function of $\chi t$ for a perfectly defined field of unit amplitude.}
\end{figure}
\begin{figure}
\begin{center}
\includegraphics[width=8cm]{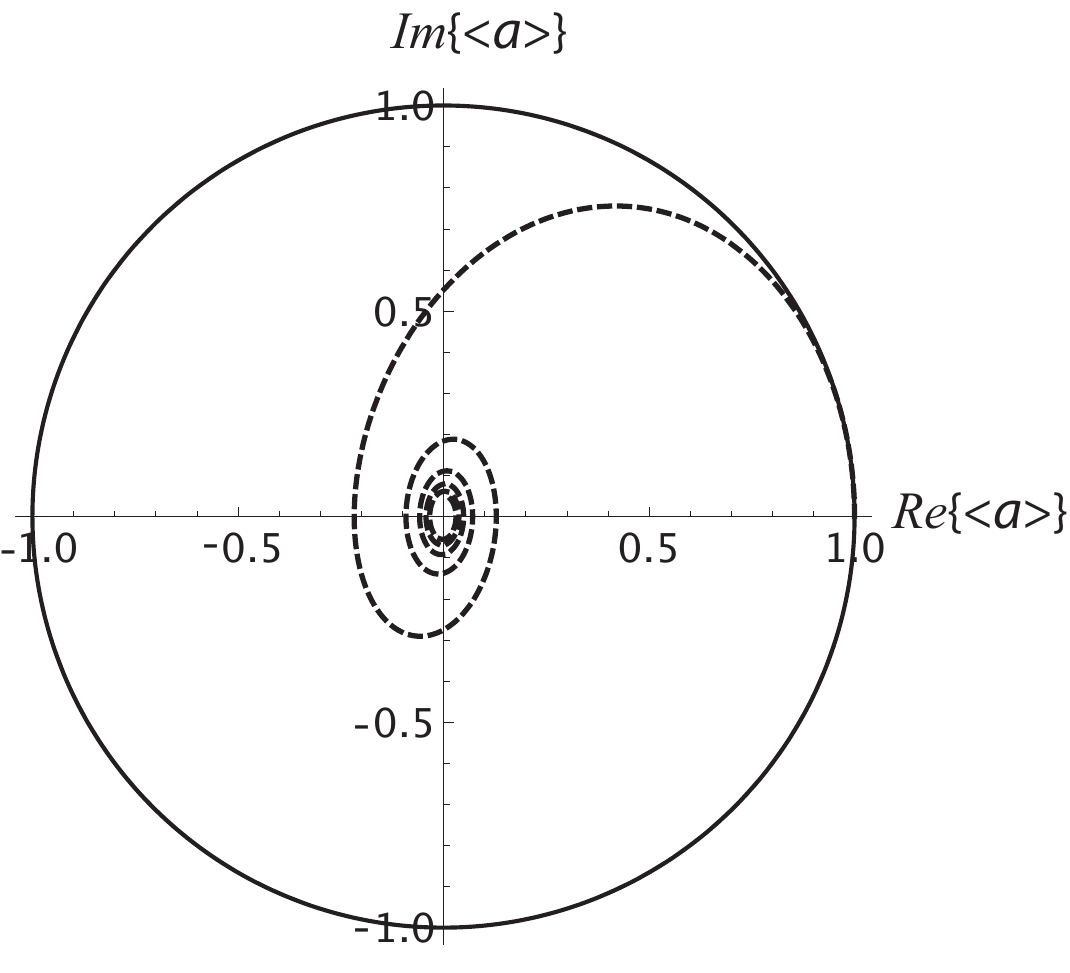}
\end{center}
\caption{Evolution of the field complex amplitude $\langle a \rangle$ as a function of $\chi t$ for a perfectly defined field of unit amplitude according to the extended semiclassical model presented in this work (dashed line) and the quantum and standard semiclassical models (solid line).}
\end{figure}

\subsection{Example: Gaussian initial field state}

For a more complete and realistic scenario let us consider a Gaussian distribution for the field in the form 
\begin{equation}
\label{6.6}
W_c (\alpha) = \frac{2}{\pi \sigma^2} e^{-2 | \alpha - \alpha_0 |^2 /\sigma^2 }   .
\end{equation}
Assuming $\alpha_0 = r_0$ real without loss of generality and using again $\alpha = r e^{- i \phi}$ we have 
\begin{equation}
W_c (\alpha) =\frac{2}{\pi \sigma^2} e^{-2 (r - r_0)^2 /\sigma^2 } e^{-(8 r r_0/\sigma^2)\sin^2 (\phi/2)}  .
\end{equation}
For $\sigma = 1$ this is the Wigner function of a Glauber coherent state, while for $\sigma < 1$ these are non quantum states, i. e., there are no legitimate  quantum states having such Wigner functions. In the extreme case $\sigma \rightarrow 0$ we get a definite complex amplitude in the form of a delta function as considered above, a field with a definite complex amplitude.

To show our point let us again focus on the phase dependence, so that the evolved field phase distribution 
becomes in this case
\begin{equation}
W_t (\phi) = \frac{1}{\pi \sigma^2}\int r \textrm{d} r \int \textrm{d}  \theta \sin \theta  \left ( 1- \sqrt{3} \cos \theta \right ) e^{ \frac{-2\left(r-r_0 \right)^2-8 rr_0}{\sigma^2} \sin^2
\frac{\phi - \sqrt{3} \chi t \cos \theta}{2}} .
\end{equation}
The result is plotted in Fig. 3 for the case $r_0=10$, $\sigma=1$, and two time instants, $\sqrt{3}\chi t =0$ and $\sqrt{3}\chi t =1$, showing clearly negativity in the evolved phase distribution. It can be also appreciated that it tends to the delta case examined above. 

\begin{figure}
\begin{center}
\includegraphics[width=8cm]{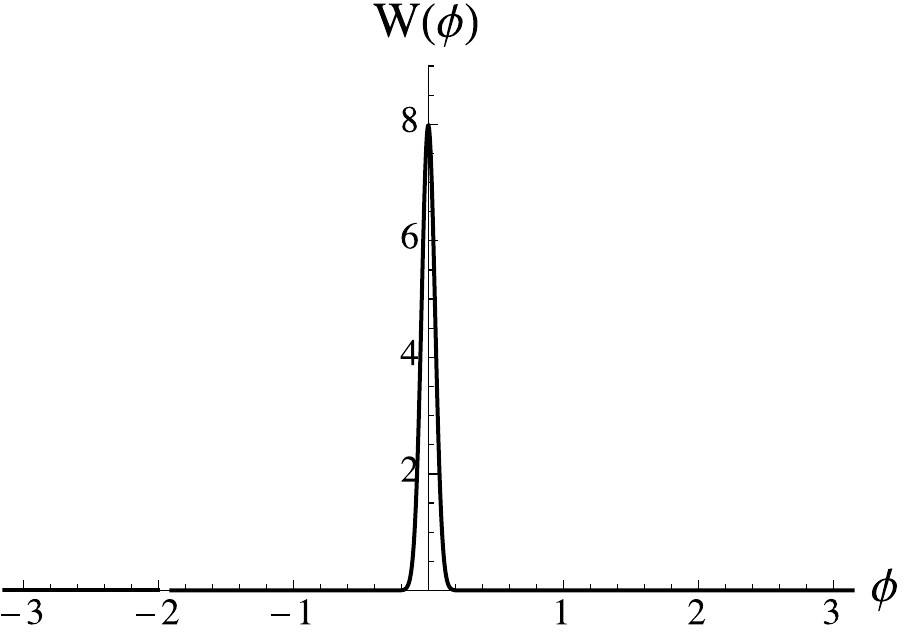}
\includegraphics[width=8cm]{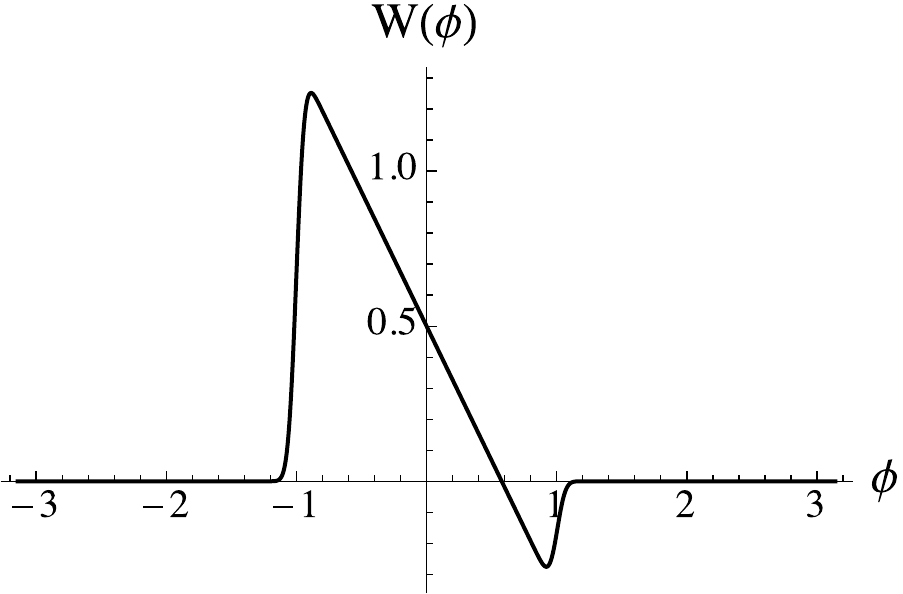}
\end{center}
\caption{Initial (upper panel) and evolved (lower panel) phase distributions for a Gaussian state  in units  $\sqrt{3} \chi t=1$,  $r_0=10$,  and $\sigma=1$. }
\end{figure}

\bigskip

Furthermore, we can express this results via the statistics of a definite observable with clear meaning both in the quantum and classical realms. Say the $Y$ quadrature $Y = i (a^\dagger-a)/2$. The statistics of $Y$ is given by the corresponding marginal of $W_t (\alpha)$ after expressing $\alpha$ as $\alpha = x + iy$ and integrating in $x$. This gives  the probability distribution $p(y)$ for $Y$ as 
\begin{equation}
p (y) =\frac{1}{\sqrt{2 \pi \sigma^2}} \int \textrm{d}  \theta \sin \theta \left ( 1- \sqrt{3} \cos \theta \right ) e^{-2 [ y + r_0 \sin( \sqrt{3} \chi t \cos \theta) ]^2/\sigma^2} .  
\end{equation}
Looking for negative values we may consider the case $\sigma=1$, $\sqrt{3} \chi t =1$, $r_0 =10$ and suitable limits of integration to get the following negative values for the integrated $p(y)$ distribution
\begin{equation}
\int_{-5}^{-1} p(y) \textrm{d} y = - 0.04 .
\end{equation}

\bigskip

\section{Atom initially in a phase state}

In the preceding example the atomic state was initially in an stationary state with a distribution around one of the poles of the Bloch sphere. Next let us consider the opposite situation of an equally weighted  coherent superposition of the excited and ground states  
\begin{equation}
| \psi_a \rangle = \frac{1}{\sqrt{2}} \left ( | e \rangle + | g \rangle \right ) , 
\end{equation}
with SU(2) Wigner function
\begin{equation}
\label{Wqps}
W_q (\Omega) = \frac{1}{4 \pi} \left ( 1+\sqrt{3} \sin \theta \cos \varphi \right ) .
\end{equation}
This can be regarded as a dipole-phase state, or as an SU(2) coherent state centred at the equator of the Bloch sphere. 

\bigskip

\subsection{Example: perfectly defined initial field state}

Following essentially the same steps of the equivalent situation in the preceding section we get for the field phase distribution 
\begin{equation}
\label{Wtphips}
W_t (\phi ) = \frac{1}{2 \sqrt{3} \chi t},
\end{equation}
for $ \sqrt{3} \chi t \geq \phi \geq - \sqrt{3} \chi t$ and $W_t (\phi) =0$ otherwise. Here again this is the effect of the back-reaction of the atom on the field phase. In this case the phase distribution is uniform and does not take negative values.

In order to compare with the other models, we compute the following mean values: 
\begin{equation}
\langle \sigma_- \rangle =  e^{-i 2 \chi r_0^2 t}, \quad  \langle a^\dagger \rangle = r_0\frac{\sin (\sqrt{3} \chi t )}{\sqrt{3}\chi t }, 
\end{equation}
while for correlations we have 
\begin{equation}
\langle \sigma_- a^\dagger \rangle = \frac{e^{-i 2 \chi r_0^2 t}}{ \chi^2 t^2} r_0 \left [\frac{  \sin (\sqrt{3} \chi t )}{\sqrt{3} \chi t} - \cos (\sqrt{3} \chi t) \right ].
\end{equation}
As in the ground state example, the correlations $\langle \sigma_- a^\dagger \rangle - \langle \sigma_- \rangle \langle a^\dagger \rangle$ tend to zero as $1/t^2$ when $t$ increases. The result for $\langle \sigma_- \rangle$ agrees with the standard semiclassical model.

\bigskip

\subsection{Example: Gaussian initial field state}

We complete the scheme analysing a classical field state with a distribution mimicking the Gaussian Wigner function of a Glauber coherent state as in Eq.  (\ref{6.6}).
The evolved phase distribution turns into:
\begin{equation}
W_t (\phi)= \frac{1}{\pi \sigma^2}\int r  \textrm{d} r \int  \textrm{d}  \theta \sin \theta  e^{ \frac{-2\left(r-r_0 \right)^2-8 rr_0}{\sigma^2} \sin^2 
\frac{\phi - \sqrt{3} \chi t \cos \theta}{2}}  .
\end{equation}
The mean values and correlations calculated follow the same structure as in the perfectly defined initial field state example:
\begin{equation}
\langle \sigma_- \rangle =  \frac{e^{-\frac{i 2\chi r_0^2t}{1+i \chi \sigma^2 t}}}{1+i \chi \sigma^2t} , \quad \langle a^\dagger \rangle =r_0\frac{\sin ( \sqrt{3}\chi t )}{\sqrt{3}\chi t},
\end{equation}
\begin{equation}
\langle \sigma_- a^\dagger \rangle = \frac{e^{-\frac{i 2\chi r_0^2t}{1+i \chi \sigma^2 t}}}{(1+i \chi \sigma^2t)^2\chi^2t^2}r_0\left( \frac{\sin \left(\sqrt{3} \chi t\right)}{\sqrt{3}\chi t}-  \cos \left(\sqrt{3} \chi t \right)\right). 
\end{equation}
As can be seen, in the limit $\sigma\rightarrow0$ we recover the mean values and correlations obtained above corresponding the case of initial field state with perfectly defined complex amplitude (see Fig. 4). \\

\begin{figure}[h]
\begin{center}
\includegraphics[width=8cm]{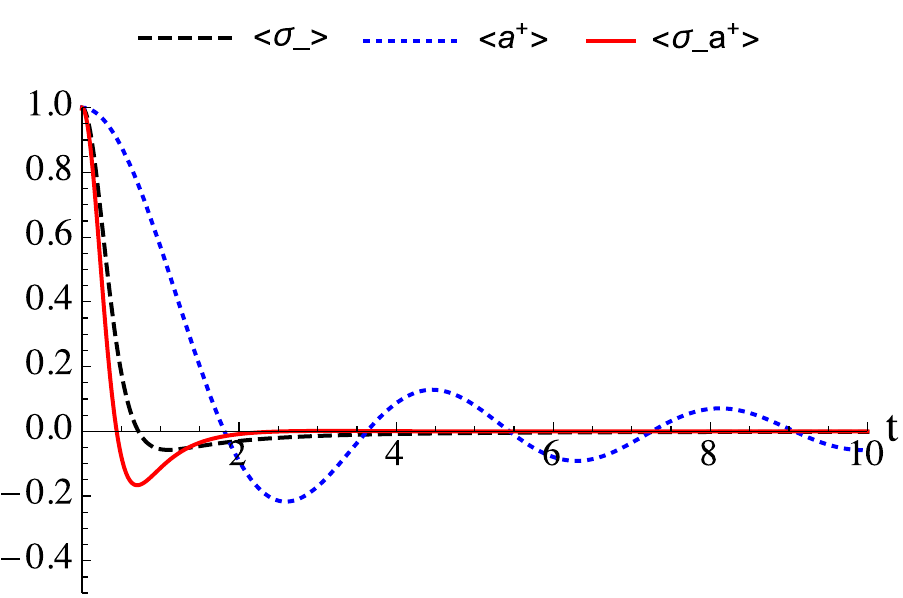}
\end{center}
\caption{Mean values $\langle \sigma_- \rangle$, $\langle a^\dagger \rangle$ and $\langle \sigma_- a^\dagger \rangle$ as functions of $\chi t$ for $\chi=1$ and a Gaussian initial field state with  $r_0=1$ and $\sigma=1$. }
\end{figure}

It is worth comparing this with the mean values for the same quantities in a fully quantum model for the field initially in a coherent state, that leads to 
\begin{equation}
\langle a \rangle = \cos (\chi t )  \langle \sigma_- a^\dagger \rangle = \frac{1}{2} e^{i \chi t} \alpha^\ast \langle \alpha e^{i \chi t}  | \alpha e^{-i \chi t} \rangle = \frac{1}{2} \alpha^\ast e^{i \chi t} e^{i |\alpha |^2 \sin (2 \chi t ) } e^{-2 |\alpha |^2 \sin^2 ( \chi t ) } . 
\end{equation}
In particular, in Fig. 5 we have plotted the correlation $\langle \sigma_- a^\dagger \rangle - \langle \sigma_- \rangle \langle a^\dagger \rangle$ as a function of $\chi t$ for $\chi=1$ and a Gaussian initial field state with  $r_0=1$,  and $\sigma=1$.

\begin{figure}[h]
\begin{center}
\includegraphics[width=8cm]{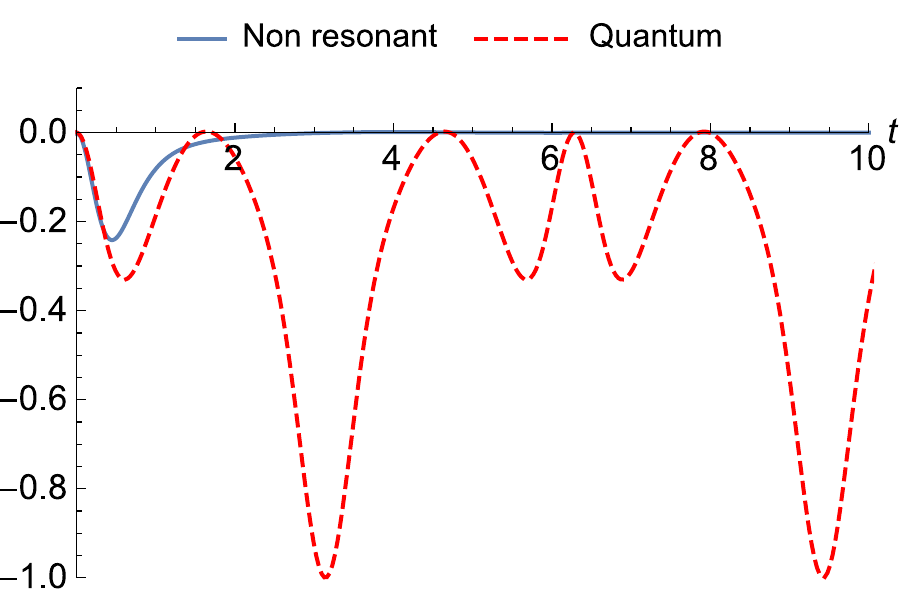}
\end{center}
\caption{Correlations $\langle \sigma_- a^\dagger \rangle - \langle \sigma_- \rangle \langle a^\dagger \rangle$ as a function of $\chi t$ for $\chi=1$ and a Gaussian initial field state with  $r_0=1$,  and $\sigma=1$.}
\end{figure}
We can appreciate that for short times there is some similarity between the extended semiclassical model and the fully quantum model. This is lost for larger times. This is natural since in the fully quantum model the evolution has a periodic character due to the discreteness of the field energy that is naturally lost in any semiclassical model. This is analogous to the collapse-revival structure of the resonant atom-field interaction. 
 
\bigskip
\bigskip

\bigskip

\section{$P$-function for classical light}

One of the advantages of the Wigner-Weyl-Moyal formulation is that it provides a highway that can be traveled in both directions. So we may ask for the field density matrix operator whose Wigner function is the one in Eq.  (\ref{Wc}). Let us dwell a little bit on the subject when the atom is in any of the two cases considered in this work and the initial field state is Gaussian with $\sigma = 1$, so $W_c$ coincides with the Wigner function of a Glauber coherent state. After  Eq. (\ref{Wc}) we can infer that the field is in a  superposition of rotated coherent states with weights given by the rotated $W_q (\Omega)$. More specifically, 
\begin{equation}
\label{rhoc}
\rho=  \int_{- \sqrt{3}  \chi t}^{\sqrt{3} \chi t} \textrm{d}  \delta P (\delta ) | \alpha_0 e^{-i\delta} \rangle \langle  \alpha_0 e^{-i\delta } | ,
\end{equation}
where $| \alpha \rangle$ are Glauber coherent states, and $P(\delta)$ is the atomic Wigner function $W_q (\theta )$ in Eqs. (\ref{Wqgs}) and (\ref{Wqps}) after the change of variables $\delta = \sqrt{3} \chi t \cos \theta$ and after integration in $\varphi$, which leads exactly to the same evolved field-phase functions in Eqs.  (\ref{pf}) and (\ref{Wtphips}). Here it comes the curious point: The atomic SU(2) Wigner function integrated in the azimuthal angle becomes the Glauber-Sudarshan $P$-function of the associated quantum state \cite{MWSZ1,MWSZ2,WC1,WC2,WC3,WC4,WC5,SU63}.

\bigskip

As another relevant feature note that the above result in Sec. VIB about negativity of the $Y$ statistics implies that $\rho$ is a nonquantum state. This is because the marginals for quadratures of the Wigner function provide their exact quantum statistics. So the above negativity of $p(y)$ implies lack of positivity of the associated $\rho$ since $p(y) = \langle y | \rho | y \rangle$ where $| y \rangle$ are the eigenstates of the quadrature $Y$.

\bigskip

\section{Conclusions}
We have presented a semiclassical model that gives the full evolution of the join atom-field system including the back reaction of the atom on the field. Our main goal is to give a non-perturbative evolution for the hybrid system composed of both quantum and classical subsystem. This approach fills the gap that exists between the standard semiclassical approaches and the full quantum theory. This gap exists because the standard semiclassical approaches neglects the back reaction of the quantum subsystem on the classical subsystem, or at most include it in terms of a mean field theory.

Therefore, the first key point of this work is to consider the nontrivial evolution of the field due to its coupling with the atom. The second key point is to use the Wigner picture of quantum mechanics, so that both systems can be properly treated alike on the same grounds. 

We have shown that the combination of these two key features transmits quantum behaviour from the quantum to the classical system. We highlight as a merit the extreme simplicity of the model that nevertheless may account for a rather complex dynamics. 

Regarding the interpretation of this result we may point two main routes that might be followed independently. On the one hand, we should expect that the mixing of two theories will necessary bring us new physics. This is to say, the distinction between classical and quantum variables is meaningless as soon as we include the full dynamics in the system.  

On the other hand, these results might serve to investigate the very consistency of the hybrid models and semiclassical calculations. From a theoretical point of view, we can not hope in general that coupling a classical system to a quantum one and keeping the distinction between both is meaningful: for a certain time it may lead to inconsistencies as such described above. Thus, even though standard semiclassical light matter interaction has been proved to be very useful to simplify the resolution of problems that quantum mechanically would be intractable, it cannot be regarded as a fundamental theory due to the facts discussed here. 

Summarizing, any fundamental theory containing on the same grounds both classical and quantum degrees of freedom, should be able to take into account dynamical mixtures of them.

\appendix

\section{Wigner-Weyl correspondences}

The Wigner-Weyl correspondence is a map between quantum operators and classical functions that serves to illustrate problems, simplify calculations, and to explore the quantum-classical borderline. Let us recall two versions. 

\subsection{Cartesian systems}

We consider an spinless (but not spineless), unbounded, Cartesian, one-dimensional system whose phase space is two-dimensional and describable by a complex variable $\beta  \propto x + i p$, its real part meaning position $x$, and its imaginary part representing linear momentum $p$. 

The correspondence holds by means of the following two relations between any operator, say $\rho$, and its Wigner functions $W(\beta)$
\begin{equation}
\label{Wr}
W(\beta) = \mathrm{tr} \left [ \rho \Delta (\beta ) \right ], \quad \rho = \pi \int \textrm{d} ^2 \beta W ( \beta ) \Delta (\beta ) ,
\end{equation}
where 
\begin{equation}
\Delta (\beta ) = \frac{1}{\pi^2} \int \textrm{d} ^2 \eta e^{\beta \eta^\ast - \beta^\ast \eta} e^{\eta b^\dagger - \eta^\ast b} ,
\end{equation}
and $b^\dagger$, $b$ are the creation and annihilation operators, say $b \propto \hat{x} + i \hat{p}$ to be more explicit.

\bigskip

Let us recall five interesting properties of this phase-space formulation not shared by other approaches, that can be extremely simply proved after the defining relations (\ref{Wr}):

\bigskip

 i) Real functions are associated to Hermitian operators and vice versa. 
 
\bigskip

 ii) The correspondence is made in both directions by just one and the same family of operators $\Delta ( \beta)$. 

\bigskip

iii) The so-called {\it traciality}, this is that quantum traces equal phase-space averages
\begin{equation}
\mathrm{tr} (A B ) = \pi \int \textrm{d} ^2 \beta W_A (\beta ) W_B (\beta ) .
\end{equation}
 
 \bigskip

iv) Classical transformation under linear transformations. This is that 
\begin{equation}
\label{We}
\mathrm{if} \quad U^\dagger \hat{z} U = M \hat{z},  \quad \mathrm{then} \quad W_{U \rho U^\dagger} ( M z )  = W_\rho ( z ),
\end{equation}
where $z$ represents all position and momentum phase-space coordinates in an arbitrary $n$-mode scenario, 
$z= (x_1,p_1,x_2,p_2,\ldots , x_n, p_n )$, $\hat{z}$ the corresponding operators, and $M$ is a $(2n) \times(2n)$ matrix.   

\bigskip

v) Proper marginals, this is that the integration of $W(\beta)$ over $p$ gives the true probability distribution for the $\hat{x}$ operator, 
and equivalently for all lineal combinations of $x$ and $p$. 

\subsection{SU(2) distributions}

We consider the phase-space representations for an angular momentum $j$ derived  from first principles in Refs. \cite{VG891,VG892,VG893}  
\begin{equation}
W (\Omega) = \textrm{tr} \left [ \rho \Lambda ( \Omega ) \right ],
\end{equation}
 with 
\begin{equation}
\Lambda ( \Omega ) = \frac{1}{\sqrt{4 \pi}} \sum_{\ell=0}^{2j} \sum_{m=-\ell}^{\ell} \sum_{k,q=-j}^{j} \sqrt{2 \ell+1}  \langle j, k; \ell, m| j, q \rangle Y_{\ell ,m} (\Omega) | j,k \rangle \langle j,q|   , 
\end{equation}
where $\langle j_1, m_1; j_2, m_2| j, m  \rangle$ are the Clebsch-Gordan coefficients,   $Y_{\ell ,m} (\Omega)$ the spherical harmonics, 
and $| j,k \rangle$ the eigenvectors of the third component of the angular momentum. Throughout $\Omega$ represents the variables on 
the phase space for the problem, this is the Bloch sphere, or Poincar\'e sphere if we refers to light polarization.  This  SU(2) distribution 
has essentially the same properties i) to iv) listed above where in such a case $U$ refers to  SU(2) transformations, that produce 
rotations on the Bloch sphere. 

\bigskip

For a two-level atom we just consider the two-dimensional case $j=1/2$
\begin{equation}
\Lambda (\Omega )= \frac{1}{4 \pi} \left ( 1 + 
\sqrt{3} \bm{\Omega} \cdot \bm{\sigma} \right ) ,
\end{equation}
where $\bm{\sigma}$ are the three Pauli matrices and 
\begin{equation}
\bm{\Omega} = \left ( \sin \theta \cos \varphi, \sin \theta \sin \varphi , \cos \theta \right ) .
\end{equation} 

The most general density matrix can be expressed as 
\begin{equation}
\label{r}
\rho = \frac{1}{2} \left ( 1 + \bm{s} \cdot \bm{\sigma} \right ) ,
\end{equation}
where $\bm{s}$ is a real vector with $|\bm{s}| \leq 1$ so that the associated Wigner function is of the form
\begin{equation}
W (\Omega )= \frac{1}{4 \pi} \left ( 1 + \sqrt{3} \bm{\Omega} \cdot \bm{s} \right ) .
\end{equation}

\section{Nonclassical and nonquantum states}

An state $\rho$ is termed nonclassical when any of its potential phase-space representatives has not the properties of a probability distribution on phase space: this is when it does not exists, it is not real, it takes negative values, or it is more singular than the delta function \cite{MWSZ1,MWSZ2}. Here we focus on the Wigner-Weyl correspondence, but any other formalism can be used for this purpose, specially the Glauber Sudarshan $P$ function \cite{MWSZ1,MWSZ2,WC1,WC2,WC3,WC4,WC5,SU63}.

The most simple example may be the case the first excited state of an harmonic oscillator  $|1\rangle$, with Wigner function 
\begin{equation}
\label{n1}
W_q (\beta) = \frac{2}{\pi} \left ( 4 | \beta |^2 - 1 \right ) e^{-2| \beta |^2} ,
\end{equation}
that  clearly takes negative values around the origin $\beta=0$.

\bigskip

We refer to a Wigner function $W(\beta)$ as nonquantum if the associated operator $\rho$ is not positive semidefinite \cite{NQ1,NQ2,NQ3}. This is the case for example of 
\begin{equation}
\label{G}
W_c (\beta) = \frac{2}{\pi \sigma^2} e^{-2 | \beta |^2 /\sigma^2 } , \quad \sigma < 1.
\end{equation}
To show this we can compute 
\begin{equation}
\label{nr}
\langle n=1 | \rho | n= 1 \rangle = \pi \int \textrm{d} ^2 \beta W_q (\beta ) W_c (\beta ) = - 2 \frac{1 -\sigma^2}{ (1 + \sigma^2)^2},
\end{equation}
which is negative for every $\sigma < 1$.

\section{Harmonic oscillator}

In this Appendix we represent the quantum system by an harmonic oscillator.  Among other possibilities this is a good approximation in many situations of light-matter interaction, as exemplified by the Lorentz oscillator model.  With this simple case, we just pretend a simple proof of principle of the main idea. The light will be represented again by a single-mode field, that is actually a perfect harmonic oscillator. 

Both systems will be represented by the corresponding complex amplitude variables: $\alpha$ for light and $\beta$ for matter. Light will be always classical and in the quantum domain, $\beta$  will be replaced by the operator $b$. In both cases the real part of these complex variables are representing coordinate and the imaginary parts linear momentum, all them in a suitable dimensionless form. Classical evolution is given by the Poisson brackets while quantum evolution is expressed within the Heisenberg picture via commutators, say in units $\hbar =1$,
\begin{equation}
\dot{\alpha} = \{ \alpha, H_c  \}, \quad \dot{b} = -i [b,  H_q], 
\end{equation}
where $H_{c,q}$ represent the corresponding Hamiltonian in classical or quantum forms. In our case 
\begin{equation}
H_c = \alpha^\ast \alpha + \mu \beta^\ast \beta + \lambda \left ( \alpha^\ast \beta + \alpha \beta^\ast \right ) ,
\end{equation}
where $\lambda$ is a coupling parameter and $\mu$ is the frequency of the quantum oscillator in units of the frequency of the classical oscillator.  Throughout we will assume perfect resonance  so that $\mu =1$. 

\bigskip

The key point is that under Hamiltonians quadratic in the $\alpha,\beta$ variables  the two following properties hold: i) The Heisenberg evolution equations are identical to the classical ones, and ii) The Wigner phase-space functions transform classically. Therefore the system evolves in phase space under the classical evolution according to the classical-like form in Eq. (\ref{We}):
\begin{equation}
W_t (\gamma (t)) = W_0 ( \gamma) \rightarrow W_t (\gamma) = W_0 ( U^{-1}  (t) \gamma) ,
\end{equation}
with 
\begin{equation}
\gamma = \pmatrix{\alpha \cr \beta}, \quad  U(t) = \pmatrix{\cos (\lambda t ) & - i \sin (\lambda t ) \cr - i \sin (\lambda t )  & \cos (\lambda t )} e^{-i t} .
\end{equation}
This places the Wigner-Weyl correspondence as an optimal arena to mix quantum and classical degrees of freedom.

Specially interesting for our purposes is that for the particular time $t= \tau$, with $\lambda \tau= \pi/2$, the quantum and classical degrees of freedom are exchanged, i. e.,
\begin{equation}
\alpha (\tau ) = - i e^{-i \tau} \beta (0) , \quad \beta (\tau ) = - i e^{-i \tau} \alpha (0) ,
\end{equation}
and then 
\begin{equation}
 W_\tau ( \alpha, \beta) =  W_c \left (  i \beta e^{i \tau} \right )  W_q \left ( i \alpha e^{i \tau} \right ) ,
\end{equation}
so the extraction of the classical and quantum parts is trivial. The key point is that, phases apart, now the distribution of the quantum degree of freedom $\beta$ is $W_c$, while for the classical degree of freedom $\alpha$ is $W_q$. Thus we get two independent proofs of the inconsistency of maintaining the distinction between the classical and quantum subsystems:

\bigskip

i) If the initial distribution for the quantum subsystem $W_q$ is nonclassical then evolution transfers nonclassicality to the assumed classical subsystem $\alpha$. For example, take the case of the first excited level $| 1 \rangle$ of the quantum harmonic oscillator in Eq. (\ref{n1})
\begin{equation}
W_q (i \alpha e^{-i \tau}) = \frac{2}{\pi} \left ( 4 | \alpha |^2 - 1 \right ) e^{-2| \alpha |^2} .
\end{equation}
Then at time $\tau$ we get for the classical system that $W_\tau (\alpha=0) <0$, that contradicts its assumed classical nature. 

\bigskip

ii) If the initial distribution for the classical subsystem $W_c$ is nonquantum,  then evolution transfers nonquantumness to the assumed quantum subsystem $\beta$. This can be the case of the Gaussian distribution in Eq. (\ref{G}) 
\begin{equation}
W_c (\alpha) = \frac{2}{\pi \sigma^2} e^{-2 | \alpha |^2 /\sigma^2 } , \quad \sigma < 1 ,
\end{equation}
that implies that the density matrix $\rho$ corresponding to $W_\tau (\beta) $ via the second relation in Eq. (\ref{Wr}) is not positive semidefinite, say $\langle  1 | \rho | 1 \rangle <0$ as shown in Eq. (\ref{nr}). This contradicts the assumed quantum nature for this subsystem. 

\bigskip

Either i) and ii) manifestly show our inability to preserve the distinction between both types of subsystems. Among other similar situations \cite{BT,CS}, it has been also proven in Ref. \cite{LG} examining up to second order moments involved in the uncertainty relations of the Heisenberg type.

\section*{Acknowledgments}

A. L. and L. A. acknowledge financial support from Spanish Ministerio de Econom\'ia y Competitividad 
Projects No. FIS2016-75199-P, and from the Comunidad Aut\'onoma de Madrid research  consortium 
QUITEMAD+ Grant No. S2013/ICE-2801.

L. A. acknowledges financial support from European Social Fund and the Spanish Ministerio de Ciencia Innovaci\'on y Universidades Contract Grant No. BES-2017-081942.

\end{document}